\documentclass[12pt,preprint]{aastex}

\slugcomment{ To appear  in ApJ, April 2011 issue}

\shorttitle{Rapid Variability in blazar S5 0716+71}
\shortauthors{chandra et al.}
\usepackage{graphicx}

\begin{document}


\title{Rapid optical variability in blazar S5 0716+71 during March 2010}


\author{Chandra S., Baliyan K.S., Ganesh S. \& Joshi U.C.}
\affil{Astronomy \& Astrophysics Division, Physical Research Laboratory,
    Ahmedabad-380009, India}

\email{sunilc@prl.res.in}

\begin{abstract}
We report  rapid optical variability for the blazar S5 0716+71 
during 2010 March 08-10 \& 19-20 in the CCD observations made from  Mt.
Abu Infrared Observatory. The light curves are constructed for the duration longer than 3-hours each night, with very high  temporal resolution($\approx$ 45 seconds in R-band). During 2010 March 08  source smoothly decayed by about 0.15 mag in 2.88 hours, 
apart from a fast flicker lasting about 30 mins. S5 0716+71
brightened up during March $ 09 $ and $ 10 $  showing high activity while it was relatively faint ($> $ 14 mag in R) albeit variable  during March 19-20. During March  9 \& 10, rapid flickers in the intensity modulate the long term intra-night ($\sim$ 3 hours) variation. The present observations suggest that the
blazar S5 0716+71 showed night-to-night and intra-night variability at various time scales with 100\% duty cycle for
variation along with microvariability at significant levels.
On night-to-night basis, the source exhibits mild bluer when brighter nature. The interaction of shocks with local inhomogeneities in the jet appears to cause intra-night variations while microvariations could be due to small scale perturbations intrinsic to the jet.

\end{abstract}


\section{Introduction}
Blazars are an extreme subclass of active galactic nuclei (AGNs), seen at small angle ( $\leq 10^\circ$) to the relativistic jet emanating from very close to the black hole \citep{Urr95}.  They are
characterized by strong variability in flux and polarization at almost all the frequencies in the electromagnetic spectrum and their variability is often used to probe the central engine and nature of physical processes in AGNs.  Blazars are known to show variations on different time scales- ranging from years to months  to days to hours and minutes.  The variations which occur during the course of night (few hours) are known as intra-night variations (INOV) while microvariability is few tenths of magnitude change in brightness during hours or less. Some authors use these terms interchangeably \citep{Wagw95, Mil89}. Fast variation, or microvariabilty,  was first discovered in 1960s by \cite{Math63}, who found  the brightness  of BL Lac object 3C48 change by  0.04 mag in V band in 15 minutes but their result was not taken seriously due to instrumental errors. However, now  
microvariabilty has been confirmed as intrinsic nature of the AGNs, especially for blazars
\citep{Mil89, Villata08,Impiom11} and has become a subject of intense activity as its physical mechanisms are not understood well.  Several models \citep[e.g.][]{Gopals91, Mw93, Marsg85, Qi91, Mar96,  Gopalw92}  have been proposed in order to explain such fast variations.  In order to constrain these models, long-term continuous observations with high temporal resolution (few minutes) are required.

The blazar S5 0716+71 (PKS 0716+714, redshift z= 0.31) 
is one of the brightest BL Lac objects which is highly variable  from the radio to $\gamma$-rays  with very high duty  cycle \citep{Wagw95}. This source has
been the target of a number of monitoring campaigns  \citep[e.g.][]{Wag96, Qi02, Rait03, Villata08}. For the first time,  INOV was detected in this source by \cite{Heidt96}. A decay in J band brightness by 0.5 mag during  2003 Dec 10-12 was reported by \cite{Baliyan05}. Variations on
various time scales have also been reported by many authors \citep[e.g.] [\& references there-in]{Gupt09, Mont06, Stalin09}. \cite{Nes02} reported 0.02 mag per hour variation during their 52 nights monitoring. Very recently, \cite{Carini11} reported B and I bands microvaribility study of S5 0716+71 based on their 5-nights observations during 2003 March 5-9. In their systematic statistical study, they detected variations at several time scales from days to few tens of minutes. \cite{Zhang08} claim microvariability down to 6 mins time scale but their sampling time is large (4 to 7 mins). In the present communication we report day to day and rapid variations in the optical brightness of BL Lac object S5 0716+71 obtained in high temporal resolution observations during the 5 nights in March 2010.

 The paper is organized as follows: Section 2 describes the observations and data analysis 
procedures while section 3 presents the  results and their discussion. Conclusions from the study are presented in Section 4.

\section{Observations \& Data Analysis}

The photometric observations were carried out by using the liquid nitrogen cooled
CCD-Camera mounted at the f/13 Cassegrain focus of the 1.2~m Telescope at Mt. Abu Infrared 
Observatory, Gurushikhar, Rajasthan, operated by the Physical Research Laboratory Ahmedabad, India. 
The PIXELLANT CCD Camera has 1296x1152 square pixels each with 22 micron size and a total 
read out time of about 13 seconds. With a scale of 0.29 arcsec per pixel, the total field 
of view is about  6.5X 5.5 arcmin$^2$. The CCD read-out noise is 4 electrons and the
dark current is negligible when cooled. The CCD-photometric system is equipped with 
Johnson-Cousin UBVRI filter set.
 
The source was observed in two observing slots; during 2010 March 08-10  and  2010 March 
19-20. All the observation nights were photometric with  a seeing better than 1''.6. Several bias
frames were taken every night at the beginning and the end of the observations. 
To construct master flats, we have taken large number of evening twilight sky flats in all 
the bands each night.
Observation strategy was to take 4-frames each in B,V, R and I band and 
then to monitor the source  in R-band for several hours. The field of view was large enough to
accommodate several standard stars in the target frame to facilitate calibration. The exposure
times were 30 secs in I, R and V bands and 60 secs in B band.

Table 1 presents the details of the observations, giving date, time (UT) of starting observation, duration of monitoring (hours) and total number of observation points on the source.
The data reduction is performed using standard routines in IRAF\footnote{IRAF is distributed by the NOAO,  operated by the Association of Universities for Research in Astronomy, Inc., under cooperative agreement with the National Science Foundation} (Image Reduction and 
Analysis Facility) software. On the bias subtracted, flat fielded images,  differential aperture photometry was performed using  DAOPHOT package available in IRAF. The photometry is carried out using several aperture 
radii, ranging from 1 to 9 times the FWHM. The right size of the aperture is 
chosen keeping in mind the optimum value of S/N ratio and the prescription of \cite{Cellone00}  to avoid spurious variations. Based on these criteria we use  4.5 arcsec as aperture radius for the target and other stars used in the differential photometry. We have used standard stars 6 (R=13.26 mag) \&  5(R=13.18 amg) from \cite{Villata98}, having apparent magnitudes close to that of the source  to check the variability of the blazar.  Such a choice of the comparison and control stars is necessary  to avoid any disparity in the measured dispersions of the target-comparison and comparison-control light curves \citep{Howell88} due to photon statistics.  One star is 
used to correct the source magnitude and the other as a control star to check the stability.

To check the significance of intra-night variability, we performed the F-test incorporated in the R statistical package.
The F-statistics is the ratio of the sample variances, or $F =  S^{2}_{B}/S^{2}_{C}$
where $S^{2}_{B}$ is the variance in the blazar magnitude and $S^{2}_{C}$
is that in the standard stars during the whole night of observations. 
For all the five nights, the F-values are more than 9 with a significance 
level of 0.999995 or $\geq 5 \sigma$. 
We have also calculated the intra-night variability amplitude 
which is given by
\begin{equation}
 Amp = \sqrt{(A_{max} - A_{min})^{2} - 2 \sigma^{2}}
\end{equation}
where $A_{max}~ \& ~ A_{min} $ are the maximum and minimum values in the light
curves and $\sigma$ is given as follows 
\begin{equation}
 \sigma = \sqrt{\frac{\Sigma (m_{i} - \bar m)^2}{N-1}}
\end{equation}
  where $m_{i}=(m_{S6}-m_{S5})_{i}$ is the differential magnitude of stars 
6 and 5  for the i-th observation point while $\bar m=(m_{S6}-m_{S5})$ is their differential magnitudes averaged
over the entire dataset, and $N$ is the number of observation points obtained that night
in a particular band.  As the errors are subtracted from the total measured variability, Eq. 1 gives fairer estimate of the amplitude of variability in the source. The results from such analysis of our data for each night are discussed in next section.

\section{Results \& Discussion}

The light curves for the source were obtained by adopting above mentioned analysis procedure.
The observed magnitudes of the source in B,V,R, \& I bands are calculated with respect to the standard star 5  and nightly averaged values are plotted as a function of time in MJD in Fig 1 for 2010 March 8-10 and 19-20. In Fig 2, we plot R-band light curves showing intra-night variations during individual nights. The bottom curve in each panel shows differential light curve of the stars. The observational uncertainties  are the $rms$ errors of the nightly differential magnitudes of the calibration star 5 and  check star 6 as given in Eq. 2.  The typical $rms$ errors  for the R-band are less than 0.008 mag.  These R-band magnitudes for S5~0716+71 for all the nights of observation are given in Table 2 in truncated form. Full data table is available from the authors.

Table 3 lists the result of the F-test, giving the date of observation, F-value,  standard deviation in the differential magnitudes of stars and the amplitude of variation in the source each night.    On March 19, we have five data points at the beginning and 24- data points at the end of the night, covering about 3.5 Hrs. The F-value for this night, therefore, is obtained from these limited measurements. The tabulated values for all the nights indicate that the source is significantly variable during present observing run, showing 100\% duty cycle for variation. Similar result has been reported by other workers \citep[eg][]{Wagw95}. The F-test results  are  in very good agreement with the values obtained from the variability test \citep{Jang97}. Here confidence level of variability is defined by the parameter  C=${\sigma_T}/{\sigma}$, where $\sigma_T$ is the standard deviation in the differential light curve of the source and comparison star. The source is considered variable at 99\% confidence level if $C \geq 2.576$. Our values for the variability parameter  for 8, 9, 10 and 20 March are, 7.66, 3.35, 4.72 and 4.45, respectively, confirming significant variability in the source on all the nights. To further support the genuine nature of the microvaribility reported here, we note that the host galaxy is more than 4-magnitude  fainter as compared to the source \citep{Nilsson08} ruling out any significant effect on the source variability.

The variation rates (mag/hour) for various peaks appearing in the light curves of each night of observation   are calculated by fitting a straight line in  the rising and falling segments using least square fitting algorithm.  
\cite{Nes02} studied intra-night variability of S5 0716+71 for 52 nights and claimed  a variation rate of 0.02 mag/hour along with a
maximum rising rate of 0.16 mag/hour while \cite{Mont06} reported equally  fast variation rates of 0.1-0.16 mag/hour.  In the following we discuss results obtained in the present study.

\subsection{Inter-night variations}

Figure 1 shows nightly averaged B, V, R \& I magnitudes as a function of time (MJD) for all 5-nights. It is evident that S5 0716+71  brightens by 0.34, 0.3, 0.3 \& 0.24 magnitudes in B, V, R and I bands, respectively,  during 2010 March 8-10. During March 19-20, source decreases in brightness by 0.23, 0.22 and 0.21 mags in B, R and I bands, respectively. Evidently, during our observations blazar S5 0716+71 was brightest on March 10 (R$\approx 12.914\pm 0.008$ mag) and faintest on March 20 (R$\approx 14.179\pm0.006$ mag). The figure also gives clear indication that the source is mildly bluer when brighter and redder when fainter. During the period of eight days (March 11-18), when we do not have observations, S5 0716+71 became fainter by about 1.10 (R-band) and 1.49 (B-band) magnitudes, clearly showing increase in the amplitude of variation with the frequency.

\subsection{Intra-night variations (INOV)} 

From the light curves in Fig 2, it is evident that the blazar  S5 0716+71 is showing significant INOV on almost all the nights it was monitored. Here we discuss the variability behaviour night by night.

On 2010 March 8, the source brightens up to $R\sim$ 13.185 mag at MJD = 55263.40. It then decays to  $R\sim$ 13.335 mag at MJD = 55263.52, a variation of 0.15 mag during 2.88 Hrs (decay rate $\approx~ 0.052$ mag/hour). This smoothly falling curve is superposed by a flicker (between MJD55263.44 and MJD55263.465) brightening the source by more than 2$\sigma$ within a time scale of about 15min.

The blazar S5 0716+71 appears to be very active during the 2010 March 9 night (cf Fig 2).  The source is in brightening phase with several rapid fluctuations modulating a smoothly varying intra-night light curve.  During the first microvariability event of the night, source fades by 0.032 mag ($> 5 \sigma$) with a time scale of $\approx$ 15 mins. It then brightens to its highest value (R=12.976 mag) at MJD=55264.456 within a time span of about 1.24 Hrs (rise rate 0.063 mag/hr). The source then decays by 0.05 mag within 0.48 hr towards the end of the observations.

 The statistical analysis of the light curves shows fast variation rates (up to $\sim$ 0.38 mag/hour) for 
several segments on March 10.
During this whole night, source remains in  bright state with rapid fluctuations in the
intensity. We have recorded most rapid fluctuation during this night. For the most 
significant peak we have calculated the rise and fall rates  $\sim$ 0.38 mag/hour  \&  $\sim$ 
0.08 mag/hour, respectively. Towards the end (from MJD 55265.502 to MJD 55265.539 ),  S5 0716+71 brightens by  0.09 mag in R-band within about 53 mins.  The  source also shows several microvariability events on a time scale of about 15 mins.
On March 19, we have few observation points in the beginning and end of the night. However, the trend shows significant variation, about 0.1 mag in 3.5 hours of duration. The light curve for this night is not shown here. On March 20, source is initially stable but later (at MJD= 55275.43) starts brightening, changing by 0.12 mag in 1.68 hrs (0.07 mag/hr). Source is generally faint during these two nights and microvariability, if any, is washed out in the relatively large scatter (0.008 mag). 
   
Let us now discuss our results in detail.
The present observations reveal significant variability at intra-night (1.24 hr to 2.88 hrs), inter-night (night-to-night to 9 nights) time scales as well as microvariability (15 mins or more) or fast fluctuations. Similar behaviour for  S5 0716+71 is reported by \cite{Carini11} in their 2003 March 5-9 observations made in B and I bands.
As mentioned above, to avoid the spurious variations caused by the variation in the seeing and/or contamination by the thermal emission from unresolved host galaxy, we have carefully chosen aperture and comparison/control stars. In order to delineate small scale fluctuations, we have used best temporal resolution ($\approx$ 45 seconds) reported so far. Many authors have reported INOV and microvariability for this source at similar time scales  \citep[e.g.][]{Querren91, Villata08, Carini11}) but with few minutes to tens of minutes temporal resolutions. \cite{Zhang08} have reported fast variations at 6 mins to 33 mins time scales with unusually large ($> 1$ magnitude) amplitude of variations. They do not mention about the aperture size adopted but have used exposures ranging from 4 mins to 7 mins and have calibrated the source magnitude with the  brightest star in the field. Some of these are prescriptions for spurious variations as investigated by \cite{Cellone00}.

 The observed optical emission in blazars originates in a part
of the accretion disk and the inner (pc-scale) regions of the jet. In the light of this, one can discuss the possible reasons
behind the variations over various time scales. We should also keep in mind that optical variability time scales shorter than a
few hours would imply emitting regions to be smaller than the Schwarzschild radius for certain objects, depending upon their mass.  There are a host of models
to explain extrinsic and intrinsic variability in blazars; microlensing effect \citep{Chang79}, light house effect
\citep{Camerzind92}, accretion disk models \citep{Chakra93, Mw93} and shock-in-jet model \citep{Marsg85}.  So far as our observations are concerned, we notice mild chromatic behaviour (bluer when brighter) in the inter-night light curves.  Our intra-night light curves do not show any symmetry or periodicity  and the variability time scales are short, ranging from few tens of minutes to few hours.  Such fast variations with  amplitude of variations reported here are difficult to explain by the accretion disk models. Since the blazar emission is dominated by the jet radiation, we concentrate on relativistic jet models.

A mild chromatic behaviour in the long term variation is explained by \cite{Villata04} and \cite{Papadakis07} for BL Lacertae, using the data obtained during several WEBT (Whole Earth Blazar Telescope; www.to.astro.it/blazars/webt/) campaigns covering the period from 1997 to 2002, by the variation in the Doppler factor due to the change in the viewing angle.  They interpreted flux variability in terms of two components; long-term (few days time scale) variation component as a mildly chromatic event and a fast (intra-day) varying component characterizing strong bluer when brighter chromatic behaviour. If the intrinsic source spectrum is well described by a power law, a Doppler factor variation does not imply a colour change.   But a mildly chromatic behaviour could be due to Doppler factor variation on a spectrum slightly deviating a power law.  A change in Doppler factor changes both, the flux ($F_\nu ~\alpha~ {\delta}^3$) and the frequency ($\nu~ \alpha ~\delta$) of emission. The inter-night variations reported here for S5 0716+71, which show mild chromatic behaviour, could be the result of the change in Doppler factor.

 Rapid variability can be produced when a relativistic shock wave or a blob propagates down the jet with
turbulent plasma  \citep{Marsgt92, Qi91}.  Synchrotron emission gets enhanced when the shock encounters particle or magnetic
field over-densities. The amplitude and the time scale of the variation depend on the turbulence and shock thickness. Fast variations,
thus require shocks to be very thin and emission to originate from very close to the central engine. Based on our shortest intra-night
variation time scale ($t_v$) of 1.24 hrs, the upper limit on the size of the emission region, with Doppler boosting and cosmological corrections, $R \leq {t_v \delta c}/(1+z)$ is
$\approx  10^{15}$ cm where $c$ is the speed of light and $\delta$ the Doppler factor (taken as 10 here).
Considering this size as a bound on the Schwarzschild radius of the central engine, one can estimate its mass using $M \approx
(c^2 R)/3G$, which comes out to be $\approx 1.1\times 10^9M_{\odot} $. However, estimation of black hole mass using such
fast optical variations must be taken with caution\citep{Querren91}. The variations shorter than INOV (microvariations) 
amounting to few tens of minutes as reported here and by many other authors can not possibly be explained  by the shock-in-jet
model. These are perhaps due either to small fluctuations intrinsic to the jet or imprinted by small fraction of black hole
horizon \citep{Begelman08}.  Such fast variations may not represent linear dimension of an emission region.

\section{Conclusions}
Here we have reported intra-night variations and microvariability in blazar S5 0716+71 as observed in the high temporal resolution observations carried out during 2010 March 8-10 \& 19-20. We note that source was variable with 100\% duty cycle at various time scales. Inter-night behaviour of the source appears to be bluer when brighter from the limited observations. It is evident that S5 0716+71 was highly active during 2010  March 9, 10 and 20 with rapid flickers superposed on the slowly varying intra-night light curves. On March 19  it 
shows substantial decay but we do not have full coverage of the night to comment on the nightly variation behaviour.

The source shows various time scales for the variation, ranging from close to two hours to 15 minutes. The inter-night variations showing mild bluer when brighter nature could be due to  variations in the Doppler factor.  While intra-night variations with few hours time scales are probably due to interaction of fast moving shocks in the jet with local small scale inhomogeneities, it is very difficult to associate faster variations with spatial extent of the emitting region. Perhaps these variations originate in a small region of the blob. Taking the shortest intra-night variability time scale of 1.24 hour, the linear size of the emitting region is estimated to be $\approx~ 10^{15}$ cm and  with corresponding Schwarzschild radius,  $\approx 1.1\times 10^9 M_{\odot}$ is mass of the black hole.

This work is supported by the Department of Space, Government of India, India.

%\bibliographystyle{natbib}
%\bibliography{sunbib} 

\clearpage

% Figures
% First Figure Fig 1
\begin{figure*}
\epsscale{0.80}
\includegraphics[scale=0.50]{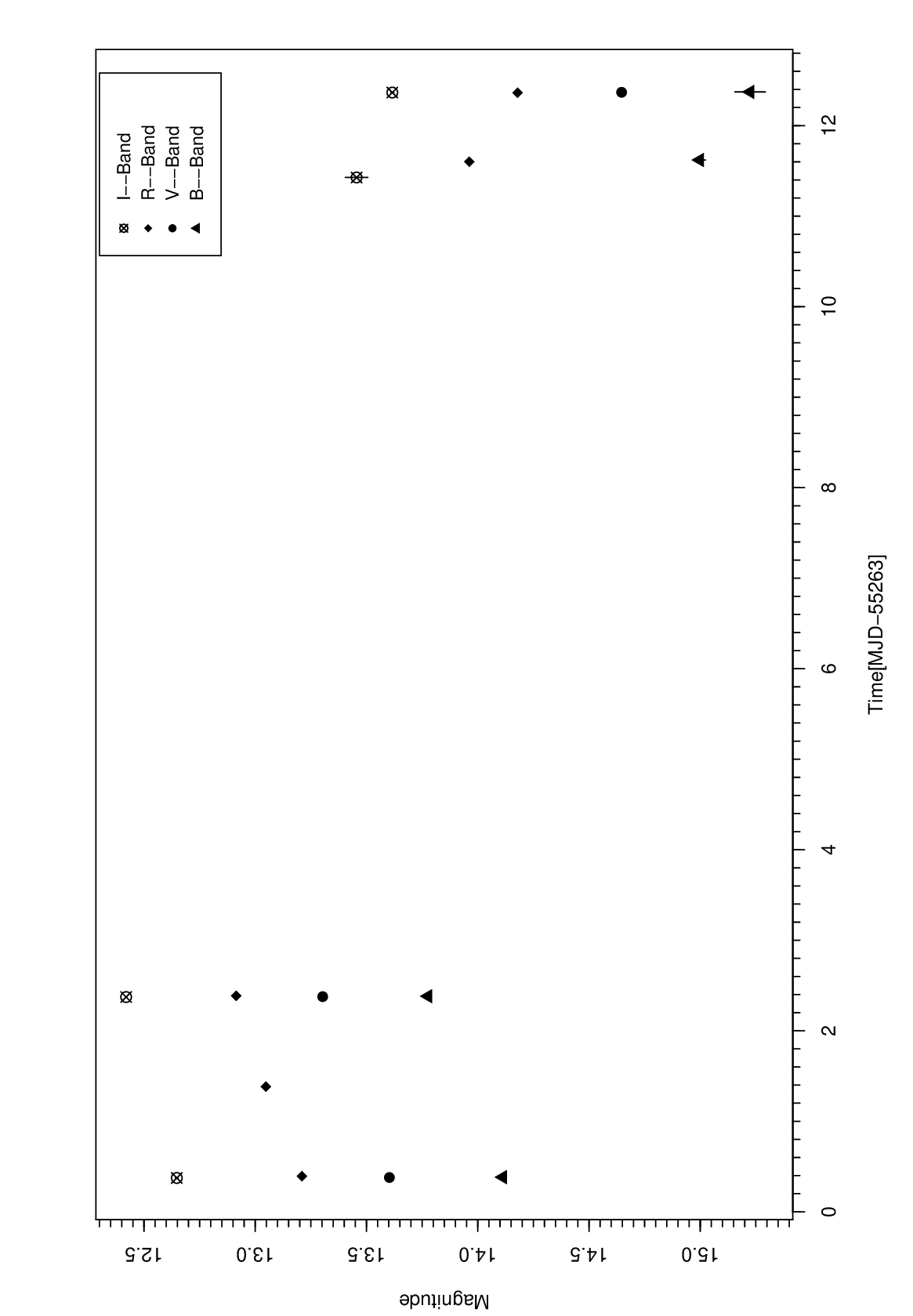}
\caption{
Nightly averaged B,V,R \& I magnitudes for S5 0716+71 as a function of time during 2010  March 8-10 \& 19-20. Most of the error bars ($\pm \sigma$) lie within the symbol.}
\end{figure*}
%  Figure Fig 2

\begin{figure*}
\includegraphics[width=5.0in]{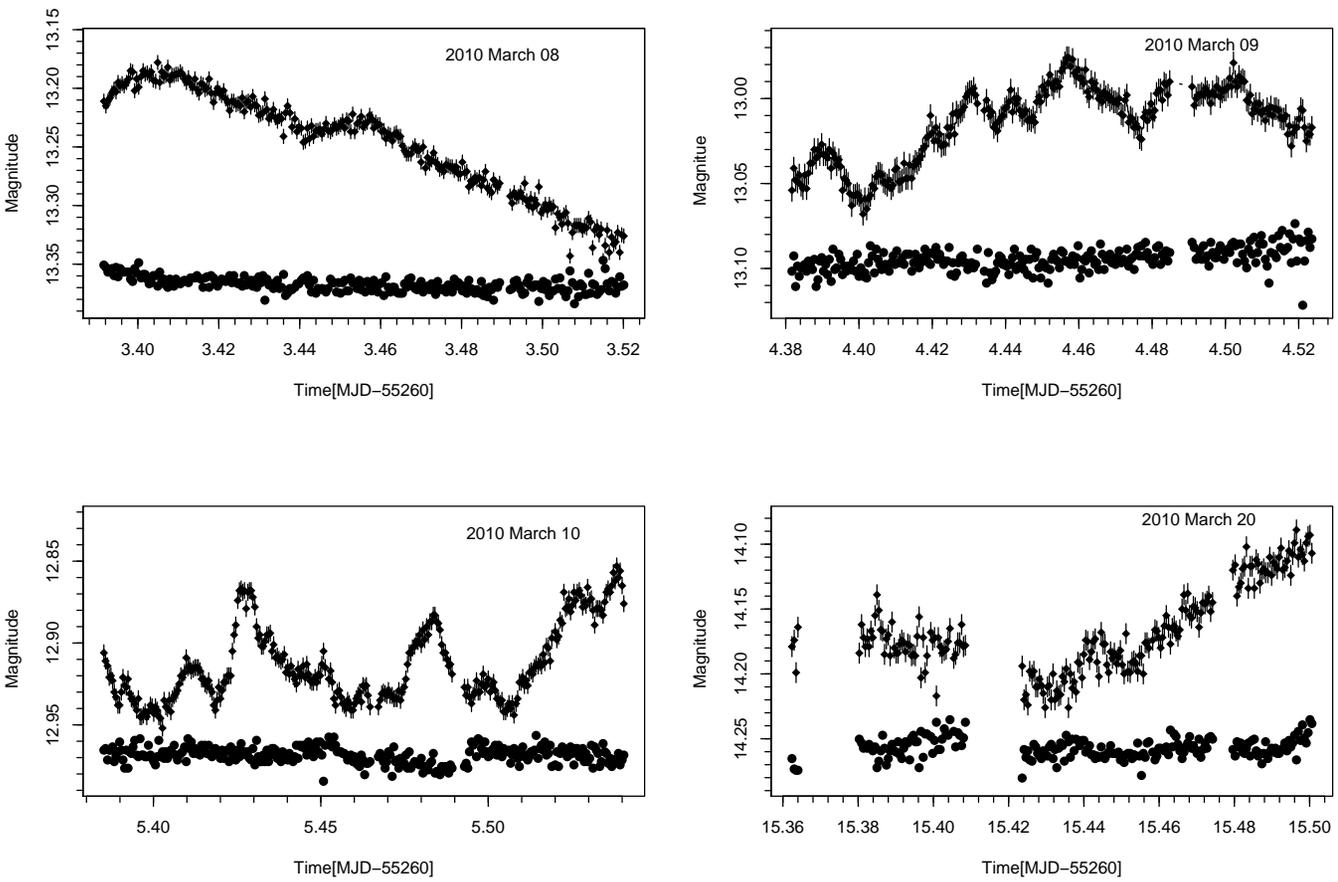}
\caption{ R-band light curves showing INOV for S5 0716+71 on 2010 March 08, 09, 10 and 20.
Lower curve in each panel shows differential light curve for comparison and control stars plotted
with appropriate offsets.}
\end{figure*}

\clearpage
% Tables

% Table 1
\begin{table}
%\begin{center}
\caption{Observation log for monitoring in R-band \label{tbl-1}}
\begin{tabular}{l c c c}
\tableline \tableline
Date & $T_{start}$(UT) & Duration ($h$) & No of images \\
%\multicolumn{1}{c}{$P$\tablenotemark{a}} & $P R_{maj}$ & $P R_{min}$ &
%\multicolumn{1}{c}{$\Theta$\tablenotemark{b}} \\
\tableline
 2010 March 08 &14:53:51 &3.2  &248 \\
 2010 March 09 &14:39:36 &3.4  &265 \\
 2010 March 10 &14:44:44 &4.0  &294 \\
 2010 March 19 &14:24:16 &0.06  &5  \\
                   &17:24:27 &0.30 &24 \\
 2010 March 20  &14:11:53 &3.3  &201 \\
\tableline
\end{tabular}
%\end{center}
\end{table}

% Table 2
\begin{table}
%\begin{center}
\caption{R-band photometric data for S5 0716+71. \label{tbl-2}}
\begin{tabular}{r c c c}
\tableline \tableline
Date & MJD (55200 +) & Mag & $\sigma$ (mag) \\
%\multicolumn{1}{c}{$P$\tablenotemark{a}} & $P R_{maj}$ & $P R_{min}$ &
%\multicolumn{1}{c}{$\Theta$\tablenotemark{b}} \\
\tableline
 2010 March 08	& 63.39160	 &   13.211 & 0.007\\
	& 63.3921	 &  13.215 & 0.007 \\
	& 63.3926	 &  13.211 & 0.007 \\
	& 63.3931	 &  13.208 & 0.007 \\
	& 63.3936	 &  13.204 & 0.007 \\
2010 March 09	& 64.381700	 & 13.054 & 0.006\\
	& 64.3822	& 13.041 & 0.006 \\
	& 64.3827	& 13.048 & 0.006\\
	& 64.3832	& 13.049 & 0.006\\
	& 64.3837	& 13.045 & 0.006\\
2010 March 10	& 65.38520	& 12.906 & 0.005\\
	& 65.3857	& 12.911 & 0.005\\
	& 65.3863	& 12.914 & 0.005\\
	& 65.3868	& 12.921 & 0.005\\
	& 65.3873	& 12.920 & 0.005\\
\tableline
\end{tabular}
\newline
\tablecomments{This table  is available in its entirety in machine readable\\
 form from the authors. Only a portion is shown representing its form and content.}
%\end{center}
\end{table}
% Table 3
\begin{table}
%\begin{center}
\caption{Intra-night variability results \label{tbl-3}}
\begin{tabular}{l c c c}
\tableline \tableline
Date & $F-Value$ &$\sigma$(mag)   & $Amp(\%)$ \\
%\multicolumn{1}{c}{$P$\tablenotemark{a}} & $P R_{maj}$ & $P R_{min}$ &
%\multicolumn{1}{c}{$\Theta$\tablenotemark{b}} \\
\tableline
 2010 March 08 &38.82 &0.006  &16.8 \\
 2010 March 09 &11.27 &0.006  &9.1 \\
 2010 March 10 &20.51 &0.005  &9.9 \\
 2010 March 19  &98.54 &0.004  &12.8 \\
 2010 March 20 &19.73 &0.008  &14.4 \\
\tableline
\end{tabular}
%\end{center}
\end{table}
\end{document}